# Revisiting the pressure-induced phase transitions of Methylammonium Lead Bromide Perovskite

Akun Liang,[1] Javier Gonzalez-Platas,[2] Robin Turnbull,[1] Catalin Popescu,[3] Ismael Fernandez-Guillen,[4] Rafael Abargues,[4] Pablo P. Boix,[4] Lan-Ting Shi,[5, 6] and Daniel Errandonea[1,*]

[1]Departamento de Física Aplicada-ICMUV-MALTA Consolider Team, Universitat de València, c/Dr. Moliner 50, 46100 Burjassot (Valencia), Spain

[2]Departmento de Física, Instituto Universitario de Estudios Avanzados en Física Atómica, Molecular y Fotónica (IUDEA) and MALTA Consolider Team, Universidad de La Laguna, Avda. Astrofísico Fco. Sánchez s/n, E-38206 La Laguna, Tenerife, Spain

[3]CELLS-ALBA Synchrotron Light Facility, Cerdanyola, 08290 Barcelona, Spain

[4]Institut de Ciència dels Materials, Universidad de Valencia, C/J. Beltran 2, 46980 Paterna, Spain

[5]Institute of High Energy Physics, Chinese Academy of Sciences (CAS), Beijing, 100049, China.

[6]Spallation Neutron Source Science Center (SNSSC), Dongguan, 523803, China

*Corresponding author: daniel.errandonea@uv.es

# ABSTRACT

The high-pressure crystal structure evolution of $CH_3NH_3PbBr_3$ ($MAPbBr_3$) perovskite has been investigated by single-crystal X-ray diffraction and synchrotron-based powder X-ray diffraction. Single-crystal X-ray diffraction reveals that the crystal structure of $MAPbBr_3$ undergoes two phase transitions following the space-group sequence: $Pm\bar{3}m \rightarrow Im\bar{3} \rightarrow Pmn2_1$. The transitions take place at around 0.8 and 1.8 GPa, respectively. This result is contradicting the previously reported phase transition sequence: $Pm\bar{3}m \rightarrow Im\bar{3} \rightarrow Pnma$. In this work the crystal structures of each of the three phases are determined from single-crystal X-ray diffraction analysis which is later supported by Rietveld refinement of powder X-ray diffraction patterns. The pressure dependence of the crystal lattice parameters and unit-cell volumes are determined from the two aforementioned techniques, as well as the bulk moduli for each phase. The bandgap behaviour of $MAPbBr_3$ has been studied up to around 4 GPa, by the means of single-crystal optical-absorption experiments. The evolution of the bandgap has been well explained using the pressure dependence of the Pb-Br bond distance and Pb-Br-Pb angles as determined from single-crystal X-ray diffraction experiments.

# I. INTRODUCTION

Metal halide perovskites form a group of materials with the simple configuration ABX$_3$ where A, B, and X are, respectively, organic parts (usually CH$_3$NH$_3^+$ (MA) or NH$_2$CH=NH$_2^+$ (FA)), metal cations, and halide anions (Cl$^-$, Br$^-$ *etc.*). Amongst these materials, MAPbBr$_3$ and MAPbI$_3$ have been found to efficiently sensitize TiO$_2$ for visible-light conversion in photoelectron chemical cells, increasing the power conversion efficiency by 3.13% and 3.81%, respectively.[1] After these results, both materials have attracted a great amount of attention. The efforts made by different research groups to study metal halide perovskites, the photovoltaic efficiency of perovskite solar cells have been soared to around 25% in 2021.[2] The tunability of bandgap energy for perovskite semiconductors is a requirement to optimize their optical properties for specific applications. For example, multi-junction perovskite solar cells, where narrow-bandgap (1.1 to 1.2 eV) and wide-bandgap (1.7 to 1.8 eV) perovskites are combined, are expected to perform with an efficiency as high as 39%.[3,4] By simply varying the ratio of I and Br in MAPb(I$_x$Br$_{1-x}$) compounds, the bandgap of hybrid perovskites can be tuned in the range of 1.6 to 2.3 eV,[5] however, this can generate instabilities due to the halide segregation.[6] Another clean method to engineer the bandgap of perovskites is by applying external pressure.[7–9] Pressure usually shortens bond distances, changing and distorting the crystal structure, and can even induce phase transitions, thereby having a significant influence on the electronic band structure.

Although several studies have been performed on the pressure-induced structural phase transitions of MAPbBr$_3$, there is still much controversy in the literature as we summarize in **Figure 1**. In 2007, Swainson *et al.*[10] investigated the pressure-induced crystal structural change of MAPbBr$_3$ with neutron diffraction up to around 3 GPa at room temperature and down to around 80 K. They reported that the crystal structure transforms from the space group $Pm\bar{3}m$ to $Im\bar{3}$, a cubic-to-cubic phase transition, at 0.87 - 1.01 GPa. They also found that MAPbBr$_3$ amorphized at around 2.8 GPa. In these experiments, 2-Propanol-d$_8$ (perdeuterated isopropanol) was used as pressure transition medium (PTM). In 2015, Wang *et al.*[7] studied the crystal structure and electronic band structure of MAPbBr$_3$ under high pressure up to 34 GPa at room temperature, by powder X-ray diffraction (PXRD) in a synchrotron light source. No PTM was used in their study. Two phase transitions were observed, from $Pm\bar{3}m$ to $Im\bar{3}$ at 0.4 GPa, and from $Im\bar{3}$ to *Pnma* at 1.8 GPa. In addition, amorphization was reported to take place at 4 GPa. The transition pressure was strongly affected by non-hydrostatic effects in this experiment. In addition, the assignment of the space group *Pnma* was not obtained by mean of indexation followed by a full-structure solution, but based on a Rietveld refinement of PXRD patterns assuming results of density-functional theory (DFT) calculations reported by Swainson *et al.*[10] However, such structure has not been experimentally found by Swainson *et al.*[10] at room temperature and high-pressure, being only observed at a temperature lower than 148 K at room pressure by using single-crystal X-ray diffraction (SCXRD).[11] On the other hand, PXRD results reported by Jaffe *et al.*[12] also contradict the existence of a high-pressure *Pnma* structure. These studies were performed using helium as PTM, which provides hydrostatic conditions up to 12 GPa.[13] In particular, Jaffe *et al*,[12] observed the phase transition from $Pm\bar{3}m$ to $Im\bar{3}$ at 0.9 GPa and the onset of amorphization at 2.7 GPa. The crystal structures of those two cubic phases were determined by SCXRD at ambient conditions and 1.7 GPa, as well as by the Rietveld refinement of the PXRD patterns from both phases. The first phase transition and amorphization pressures are consistent with that report by Swainson *et al*[10] Four other high-pressure studies can be found in the literature. Kong *et al.* carried out studies only up to 1 GPa.[14] They only reported the phase transition from $Pm\bar{3}m$ to $Im\bar{3}$ at around 0.5 GPa. In their case, silicone oil was the PTM. On the other hand, the first phase transition was observed at 0.75 GPa by Szafrański *et al.*[8] In their work the $Im\bar{3}$ phase coexisted with an unknown phase (named Phase VII in their work) in the pressure range 2.1-2.7 GPa. These authors correlate changes in the crystal structure with changes in the bandgap. They propose that the bandgap energy of MAPbBr$_3$ may have a linear relationship with the Pb-Br bond length. The pressure-induced crystal structure phase transition has been also investigated by Zhang *et al.* with three different quasi-hydrostatic conditions (Helium, Argon, and no PTM).[15] In the experiment where helium was used as the PTM, SCXRD was used to characterize the crystal structure, the pressure-induced phase transition from space group $Pm\bar{3}m$ to $Im\bar{3}$ have been observed at 0.85 GPa, followed by an isostructural phase transition at 2.7 GPa, which was accompanied by a unit cell volume collapse of around 4.4 Å$^3$. In the second experiment where argon was used as PTM, the first phase transition was observed at 1 GPa, after that the $Im\bar{3}$ phase coexisted with the *Pnma* phase up to the highest pressure in their work (11.9 GPa). The reason used to justify the phase coexistence was the solidification of argon at around 1.4 GPa and room temperature.[13]

In the experiment where no PTM was used, the first phase transition was observed at the lowest pressure of 0.4 GPa, in agreement with the transition pressure reported in the work from Wang *et al*.[7] followed by another pressure-induced phase transition from $Im\bar{3}$ to *Pnma* was found at 1.5 GPa. Finally, in the PXRD experiment of Yin *et al*,[16] the first phase transition ($Pm\bar{3}m$ to $Im\bar{3}$), second phase transition ($Pm\bar{3}m$ to *Pnma*) and amorphization were located at 0.99 GPa, 2.41 GPa and 4.06 GPa, respectively, however the use of PTM is not reported in this work.

In summary, the use of different pressure transmitting media could affect the structural sequence and phase transition pressures observed in MAPbBr$_3$. The crystal structures of the two low pressure cubic phases have been unambiguously characterized by neutron diffraction, PXRD, and SCXRD and there is an agreement about this fact in the literature. However, the second pressure-induced phase transition, from space group $Im\bar{3}$ to *Pnma*, has only been observed in three papers,[7,15,16] and the crystal structure used to perform Rietveld refinements on PXRD patterns was not properly solved because it was adopted from the phase observed at low temperature and ambient pressure. The pressure-induced amorphization was indeed observed in most of the reported papers, but at different pressures.

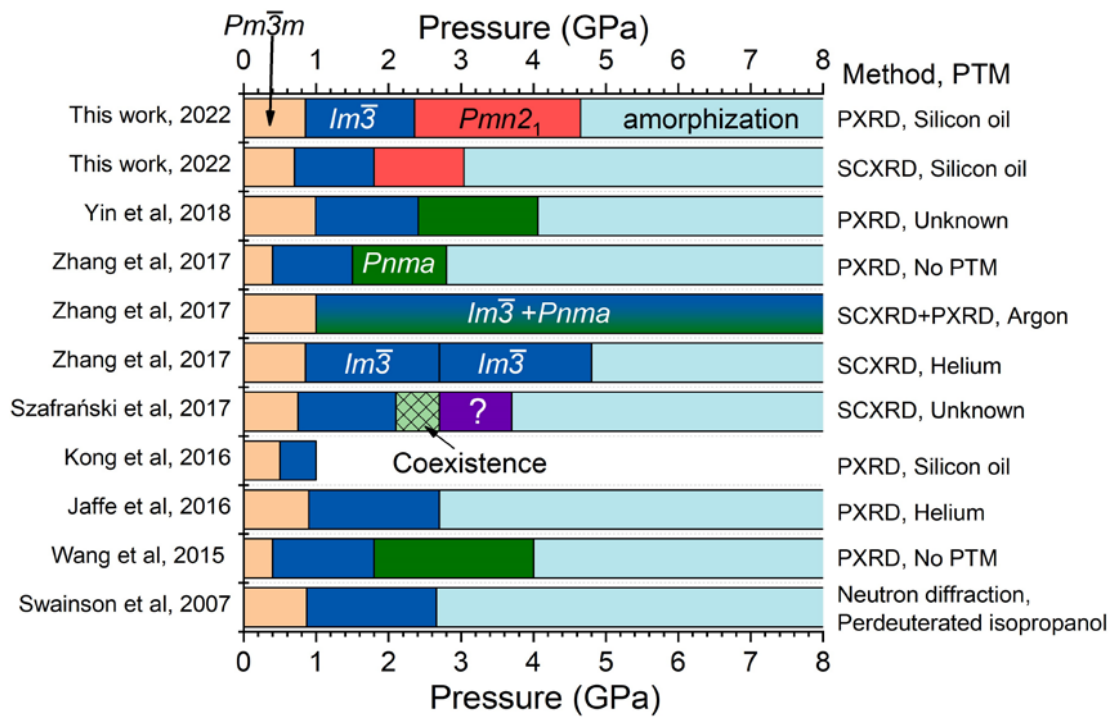

Figure. 1. Summary of the pressure-induced phase transitions observed in MAPbBr$_3$ reported in the literature. Including the results reported by Swainson *et al*,[10] Wang *et al*,[7] Jaffe *et al*,[12] Kong *et al*,[14] Szafrański *et al*,[8] Zhang *et al*,[15] Yin *et al*,[16] and this work. The different crystal structures with different space groups are shown in different colors. The diffraction method, single-crystal X-ray diffraction (SCXRD), powder X-ray diffraction (PXRD) and pressure transition medium (PTM)

used in the studies are shown on the right-hand side.

In this work, the pressure-induced crystal structure phase transitions of MAPbBr$_3$ have been re-examined by SCXRD up to 5 GPa. The crystal structures have been well established up to 3 GPa, the pressure-induced phase transitions have been further confirmed by the Rietveld refinement on PXRD phases and changes in bandgap energies have been investigated and explained. Two pressure-induced crystal structure phase transitions were founded both in SCXRD, PXRD, and optical experiments. The detailed crystal structure information of the three phases obtained from SCXRD will be reported, as well as the pressure-induced change in the lattice parameters and equations of state. The pressure-induced amorphization has been also observed in both SCXRD and PXRD.

## II. METHODS

### A. Sample preparation

Lead bromide (PbBr$_2$, 98% purity, purchased from Fisher Chemical), Methylammonium bromide (MABr, 98% purity, purchased from Ossila), and Dimethylformamide (DMF, 99.8% purity, purchased from Sigma Aldrich) were used as the starting materials. Lead bromide and methylammonium bromide were dissolved in DMF (1 M). The solution was stirred at ambient conditions until the precursors were completely dissolved. The solution was then filtered with 0.2 mm pore size filter, kept in a closed vial of 20 cm$^3$, and heated up to 80 ºC in an oil bath. The temperature ramp was set to 20 ºC/h until 60 ºC. Then, the solution is heated until 80 ºC with a temperature ramp of 10 ºC/h. Finally, it was kept at 80 ºC for 24 hours. Reproducible size crystals are obtained with this method. The fine powder sample was obtained from grounding the single-crystal sample.

### B. X-ray diffraction

#### 1. High-pressure single-crystal X-ray diffraction

SCXRD has advantages over the PXRD approach because it decouples the fitting of lattice and structural parameters, leading thus to higher resolution. In this study, SCXRD measurements were performed at room temperature using a Rigaku SuperNOVA diffractometer equipped with an EOS charge-coupled device (CCD) detector and a molybdenum radiation micro-source ($\lambda = 0.71073$ Å). All measurements were processed with the CrysAlisPro software.[17] Numerical absorption corrections based on Gaussian integration over a multifaceted crystal model were applied using the ABSORB-7 program.[18] For HP measurements, a Mini-Bragg diamond-anvil cell with an opening angle of 85° and anvil culets of 500 μm diameter was used to generate the high-pressure environment. A stainless-steel gasket with a centered hole of 200 μm diameter and 75 μm depth was used as the gasket. Silicone oil was used as pressure-transmitting medium (PTM).[13] The sample was placed on one of the diamond anvils (diffracting side), together with a small ruby sphere used as a pressure sensor.[19] The crystal structure was refined for each pressure, using previous results as starting points, against $F^2$ by full-matrix least-squares refinement implemented in the SHELXL program.[20]

## 2. High-pressure powder X-ray diffraction

*In situ* PXRD experiments were performed at the BL04-MSPD beamline of ALBA-CELLS synchrotron.[21] A membrane LeToullec-type diamond anvil cell (DAC), with a culet of 400 μm in diameter, was used to generate the high-pressure environment. A hole with a diameter of 200 μm drilled in the center of a pre-indented stainless-steel gasket served as the sample chamber. As in SCXRD experiments, silicone oil was used as the PTM, and the ruby fluorescence method was used for pressure determination.[19] The wavelength of the monochromatic X-ray beam was 0.4642 Å, and the spot size of the X-ray was $20 \times 20$ μm (full width at half maximum). A Rayonix SX165 CCD image plate was used to collect the diffraction patterns, and the sample-to-detector distance was calibrated using a $LaB_6$ standard. The collected two-dimensional diffraction images were reduced to conventional XRD patterns using DIOPTAS.[22] The FullProf[23] suit was used to perform Rietveld refinements.[24]

## C. High-pressure optical absorption

A membrane-type DAC was used to generate the high-pressure environment, the culet of the diamond was 400 μm. A stainless-steel gasket was first pre-indented to a thickness of 40 μm, then a 200 μm in diameter hole was drilled in the center and served as sample chamber. A single-crystal sample, together with silicone oil (PTM) and a ruby sphere (pressure gauge), were loaded in the sample chamber. The sample-in and sample-out method was used to acquire the optical absorption spectra in a home-built optical setup, consisting of a tungsten lamp, fused silica lenses, reflecting optics objectives (15×), and a visible-near infrared spectrometer (Ocean Optics Maya 2000 pro). The light transmitted through the sample [$I(\omega)$] was normalized by the intensity of the light transmitted through the PTM [$I(\omega_0)$]. More details on the experimental set up can be found in our previous work.[25–27]

## III. RESULTS AND DISCUSSION

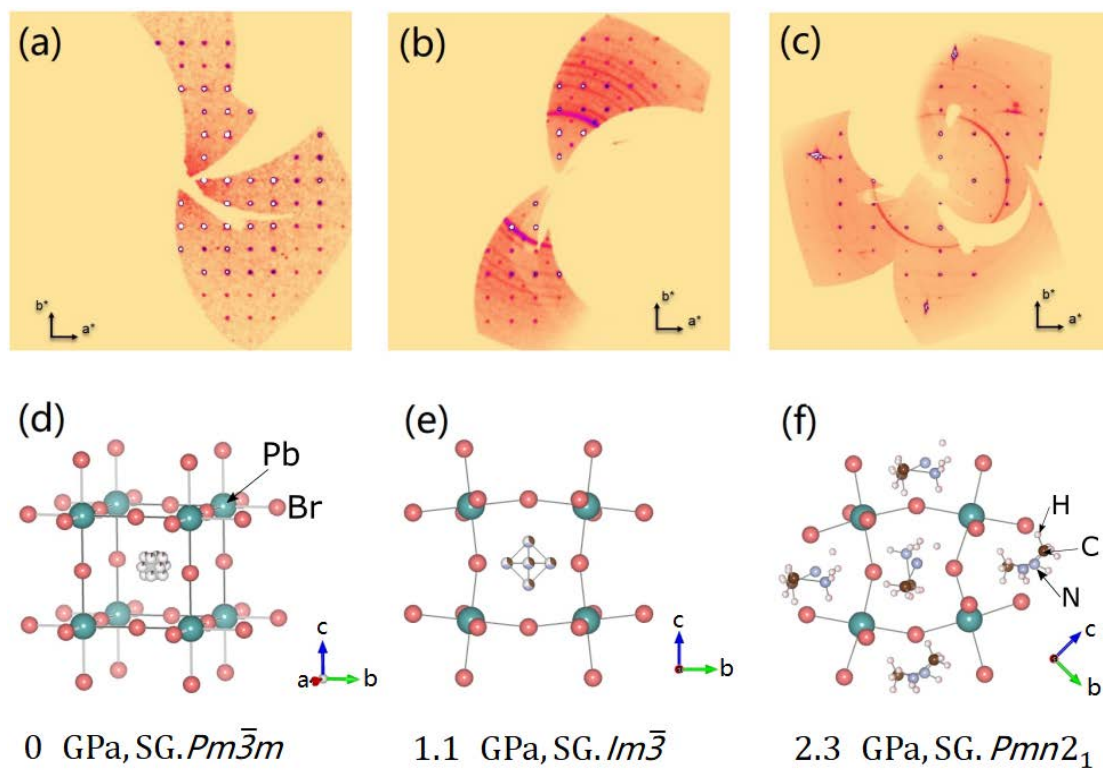

Figure. 2. Results of high pressure SCXRD experiments on MAPbBr$_3$. (Top) Reconstructed reciprocal-space precession-images for the (*hk*0) plane at (a) ambient conditions, (b) 1.1 GPa, and (c) 2.3 GPa. They correspond to phases I, II, and III. In phases I and II the order of rotational symmetry is 4. In phase III it is 2. (Bottom) (d)-(e) show the crystal structure of MAPbBr$_3$ obtained from the SCXRD data shown above. The space group (SG) of each crystal structure is shown at the bottom. The atoms are shown in different colors as indicated in the figure.

SCXRD images at different pressures are shown in **Figure 2**, as well as the crystal structures obtained from the experiments. Details of the data collections, refinement results and quality factors, detailed atomic positions, and crystal structure information at the three different pressures can be found in **Tables S1-S4** of **Supplementary Information**. CIF files can be obtained from Cambridge Crystallographic Data Centre (deposit numbers 2194528-30). At ambient conditions, MAPbBr$_3$ crystallized in the cubic structure, described by the space group $Pm\overline{3}m$. The crystal structure determined here at ambient conditions is in agreement with that reported in all the previous works,[7,8,10,12,14–16] and it schematically represented in **Figure 2**. Here we name it as Phase I. In this structure Pb atoms are bonded with six Br atoms forming a regular octahedron. The six Pb-Br bonds have a length of 2.9642 ± 0.0011 Å. The PbBr$_6$ octahedra are bridged by corner sharing Br atoms. The Pb-Br-Pb angle is 180º forming PbBr$_6$ octahedra in a linear chain. The organic molecule is located at the center of the cubic structure with an important positional disorder. The SCXRD pattern collected at 1.1 GPa is different from that collected at ambient conditions (**Figure 2b**), we also found change in the PXRD experiment at a similar pressure as we show later in this section. Both results support that a pressure-induced structural phase transition has taken place. Here we name the second phase as Phase II, the phase transition pressure (Phase I to Phase II) we found in SCXRD experiment is 0.81 GPa, in agreement with the phase transitions pressure reported in Ref.[8,10,12,15,16] where Helium or perdeuterated isopropanol were used as PTM (**Figure 1**). The crystal structure of Phase II determined from our SCXRD data can be described by the space group $Im\overline{3}$. It is consistent with the results reported in Ref.[7,8,10,12,14–16] determined from neutron diffraction, SCXRD or PXRD. In Phase II (**Figure 2e**), the PbBr$_6$ octahedron remain regular, the six Pb-Br bonds are equal in length, and the bond distance is 2.9304 ± 0.0012 Å at 1.1 GPa. However, the Pb-Br-Pb bonds are not straight anymore, the angle of Pb-Br-Pb is 161.4 ± 0.3 °. A second pressure-induced phase transition was observed at 1.8 GPa in our SCXRD experiment, diffraction data at 2.3 GPa are shown in **Figure 2c**. The crystal structure determined is orthorhombic (**Figure 2f**) and the space group is $Pmn2_1$ (No. 31). Here we name the third phase as Phase III. The crystal structure determined here from SCXRD is different from the previous results (space group *Pnma*),[7,15,16] and it is confirmed by the Rietveld refinements of our PXRD patterns as shown later. For the crystal structure of Phase III collected at 2.3 GPa (**Figure 2f**), the PbBr$_6$ octahedra are not regular anymore, Pb are located at two different Wyckoff positions and Br are located at eight different Wyckoff positions. The Pb-Br bond distances range from

2.859 ± 0.015 Å to 3.034 ± 0.015 Å, wherein the Pb-Br-Pb angle varies in the range of 142.0 ± 0.5 to 172.1 ± 0.5 degrees. At pressures higher than 3 GPa in the SCXRD experiment, the quality of the diffraction data quickly decreases, probably due to degradation of the monocrystal, and it becomes impossible to resolve the structure. A possible reason for this is the introduction of disorder in the crystal structure as a precursor of the pressure-induced amorphization of the sample.[28]

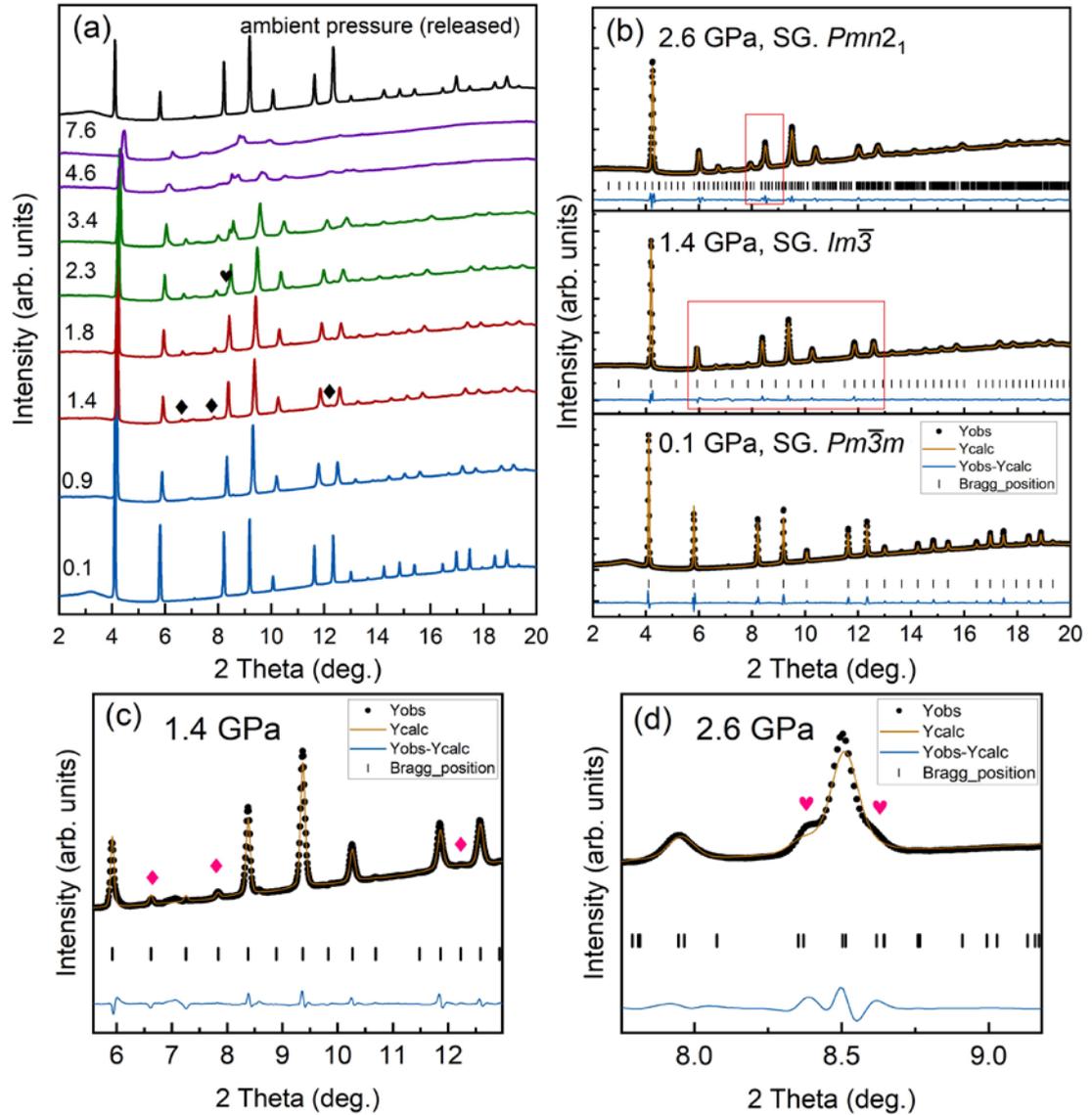

Figure. 3. Results of high pressure PXRD experiments on MAPbBr$_3$. (a) PXRD patterns at selected pressures, patterns from different phases are shown in different colors. Phases I, II, and III are shown in blue, red, and green, respectively. The patterns in purple show signs of a pressure-induced gradual of loss long-range order. Pressures are given in GPa. The black diamond and heart symbols identify the appearance of new reflections. (b) Typical Rietveld refinement at 0.1 GPa (Phase I), 1.4 GPa (Phase II), and 2.6 GPa (Phase III). (c) and (d) Enlarged images of the areas marked by red boxes in (b) for experiments collected at 1.4 GPa and 2.6 GPa, respectively. The black dots are experimental results (Yobs), the refined patterns (Ycalc) are shown in solid yellow lines, the difference between experiments and refinements (Yobs-Ycalc) are shown in solid blue lines. The vertical ticks show the position of diffraction peaks (Bragg_position).

PXRD patterns of MAPbBr$_3$ at selected pressures are shown in **Figure 3a**. At pressures lower than 0.9 GPa, they can be well refined by the ambient-pressure cubic crystal structure (Phase I, space group: $Pm\bar{3}m$, $R_p$=1.05 and $R_{wp}$=1.97) obtained from SCXRD experiment. As an example, we provide in **Figure 3b** the Rietveld refinement at 0.1 GPa. At 1.4 GPa, there are two additional peaks located between 6 and 8 degrees (marked by black diamonds in **Figure 3a** and pink diamonds in **Figure 3c**). Notably, the same extra peaks also have been observed in the PXRD patterns reported in Ref.[7,15,16]. Another additional peak can be observed at 1.4 GPa at around 12 degrees. This peak is too weak to be observed in **Figure 3a**, but it can be identified in **Figure 3c**. The emergence of the new peaks indicates a pressure-induced phase transition. Furthermore, the Rietveld refinement of the PXRD pattern at 1.4 GPa (**Figures 3b and 3c**) shows that all peaks can be explained by the cubic crystal structure described by the space group $Im\bar{3}$ as we determined for the Phase II in SCXRD experiment ($R_p$=0.60 and $R_{wp}$=1.17). At pressures above 1.8 GPa, two extra peaks appear on either side of the peak located at around 8.2 degrees (marked by a black heart in **Figure 3a** and pink hearts in **Figure 3d**), indicating another pressure-induced phase transition. These extra peaks have also been observed in Ref.[16], but in that work the space group of the third phase has been assigned to *Pnma* following the assignment made in Ref.[7]. However, the structural solution made in Ref.[7] raises some doubt since there is a peak at low angles not explained by the proposed structure and there are peaks predicted by the *Pnma* structure not observed in the experiments. In contrast, as the Rietveld refinement of the PXRD collected at 2.6 GPa (**Figures 3b** and **3d**) shows ($R_p$=0.61 and $R_{wp}$=1.13), the PXRD pattern can be satisfactorily explained by the crystal structure with a space group *Pmn*2$_1$ (No. 31) determined from our SCXRD experiments for Phase III. In all our experiments, the new peaks (which are the sign of pressure-induced phase transitions, both Phase I → Phase II and Phase II → Phase III) are properly indexed by the crystal structure determined from SCXRD in this work (**Figures 3c** and **3d**). Therefore, the Rietveld refinements of the PXRD patterns confirm the crystal structure determined from SCXRD. With increasing pressure beyond 3.4 GPa, the intensity of the diffraction peaks is reduced, most peaks become broader and most peaks for values of 2$\theta$ higher than 11° disappear (as the purple PXRD patterns show in **Figure 3a**). This might be caused by a gradual disordering of the crystal structure related with the partial amorphization of MAPbBr$_3$ which was proposed to occur based on previous PXRD experiments.[7,12,15,16] The most likely picture is the disorder of MA within an ordered inorganic PbBr$_6$ framework which is preserved. Furthermore, the pressure-induced structural changes in MAPbBr$_3$ are totally reversible, as shown by the PXRD pattern collected after the pressure was released to ambient pressure (see the topmost spectra in **Figure 3a**). This is consistent with the reversibility found in works.[7,15,16]

The lattice parameter and the unit-cell volume per formula unit as a function of pressure obtained from our experiments are plotted in **Figure 4** and the data can be found in **Tables S5** and **S6** in the **Supplementary Information**. From the structure information of the three phases summarized in **Table. S1** in the **Supplementary Information**, it can be seen that lattice parameters from Phases II and III nearly doubled the lattice parameter from phase I, and consequently the unit-cell volume becomes approximately 8 times that of Phase I. Then, for a better comparison in **Figure 4**, the lattice parameters from Phases II and III are divided by 2 and the unit-cell volume per formula is represented. There is no observable discontinuity in the unit-cell volume at the phase transitions (**Figure 4b**). On the other hand, at the second transition the crystal structure is elongated in one direction (*b* axis) and shortened in other (*c* axis), while the third direction remains unmodified (*a* axis). In the figure it is shown that the lattice parameters obtained from PXRD and SCXRD show a good agreement with each other.

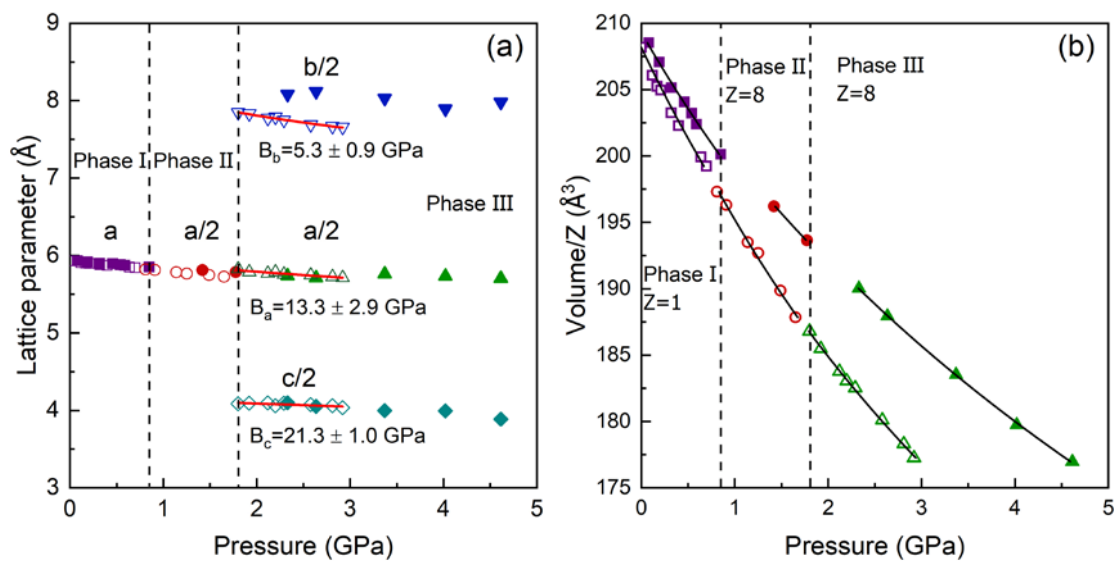

Figure. 4. Pressure dependence of the lattice parameters and unit-cell volume of MAPbBr$_3$. (a). Crystal lattice parameters obtained from PXRD (solid symbols) and SCXRD (empty symbols) as a function of pressure. The lattice parameters in Phase II and III have been divided by 2 to better

compare with Phase I. The vertical dashed lines indicate the phase transition pressure by considering the data form both PXRD and SCXRD experiments. The red solid lines in Phase III are the EOS fitting of the lattice parameter obtained from SCXRD experiments. (b). Unit-cell volume per formula unit as a function of pressure obtained from PXRD (solid symbols) and SCXRD (empty symbols). The black solid lines are the second-order Birch-Murnaghan fitting.

The unit-cell volumes per formula unit for each phase are have been fitted separately by second-order Birch-Murnaghan equations of state (BM-EOS) (**Figure 4b**).[29,30] The obtained bulk modulus and its pressure derivatives are summarized in **Table I**, together with the value reported in the previous works by using different experiment methods and PTM. In this work, the bulk moduli for Phases I and II obtained by fitting the unit-cell volume per formula unit (according to SCXRD) as functions of pressure, are in agreement with the values reported in Ref.[10], where neutron diffraction was used to measure the unit-cell volume, and isopropanol was used as PTM, also in agreement with the value reported in Ref.[15], in which the unit-cell volume of $MAPbBr_3$ is determined from SCXRD experiment with helium as PTM. However, we did not find any sign of the pressure-induced isostructural alleged phase transition at around 2.7 GPa in Ref.[15] and the accompanied ~4.4 $Å^3$ drop in the volume. The bulk modulus for Phase I from the PXRD experiment in this work is similar to the data reported in Ref.[12], where the probing method is PXRD and helium was used as the PTM. Unfortunately, the bulk modulus in phase II obtained from the PXRD in this work only contain two data, and it is higher than any reported values in the literature. There is no reported experimental bulk modulus of Phase III, which is described by space group *Pnma* in previous work[7,15,16] but unambiguously by *Pmn*$2_1$ in this work. It is 13.0 GPa and 19.1 GPa calculated from the SCXRD and PXRD experiment, respectively. There is a discrepancy in the bulk modulus determined From SCXRD and PXRD in this work, the same phenomenon has also been observed in previous works which show a larger bulk modulus in PXRD than that determined from SCXRD experiment when using helium as PTM.[12,15] Therefore, the differences cannot be related to deviatoric stresses induced by non-hydrostaticity. Similar differences have been observed in other compounds, Like $FeVO_4$, $PbCrO_4$, and $BiMnO_3$[31–33] being related to the existence of grain-grain stresses in powder XRD experiments. It should be noted here that the three phases are highly compressible with values of the bulk modulus comparable to that of metal-organic frameworks.[34]

We also fitted the lattice parameters *a*, *b* and *c* obtained from our SCXRD experiment (**Figure 4a**) in Phase III with a similar modified Birch's EOS[35] as following:

$$\frac{a}{a_0} = (1 + \frac{B'_a}{B_a}P)^{-1/3B'_a}$$

where $a_0$ is the lattice parameter of MAPbBr$_3$ at normal conditions, $B_a$ is the linear moduli in $a$ axis, $B_a$' is the pressure derivative. As in the second-order BM-EOS, here we fixed $B_a$' in a value of 4. The ambient-condition lattice parameter and moduli are obtained from the fitting. The moduli of the crystal structure in axis $a$, $b$ and $c$ are 13.3 ± 2.93 GPa, 5.3 ± 0.9 GPa and 21.3 ± 1.0 GPa, respectively. The crystal structure in Phase III shows an anisotropic behavior under compression and $b$ axis is the most compressible axis.

TABLE I. Summary of the bulk moduli ($B_0$) for different phases of MAPbBr$_3$. "ND" means neutron diffraction, "PXRD" powder X-ray diffraction, "SCXRD" single-crystal X-ray diffraction and "DFT" computer simulations using density-functional theory. Other information, like pressure transition medium (PTM) used in experiment, the order of Birch-Murnaghan equation of state (BM-EOS), zero-pressure volume per formula ($V_0$/Z) are also included in this table.

| Phase | Method | PTM | BM-EOS | $V_0$/Z (Å$^3$) | $B_0$ (GPa) | $B_0$' | Ref. |
|---|---|---|---|---|---|---|---|
| $Pm\bar{3}m$ | ND | isopropanol | 2$^{nd}$-order | 208.1 (1) | 15.6 (4) | 4.0 | 10 |
| | PXRD | Helium | 2$^{nd}$-order | 207.8 (5) | 17.6 (4) | 4.0 | 12 |
| | SCXRD | Helium | 2$^{nd}$-order | ~208 | 12.2 (8) | 4.0 | 15 |
| | SCXRD | Silicone oil | 2$^{nd}$-order | 208.2 (1) | 14.0 (3) | 4.0 | a |
| | PXRD | Silicone oil | 2$^{nd}$-order | 208.5 (1) | 19.6 (8) | 4.0 | a |
| $Im\bar{3}$ | ND | isopropanol | 2$^{nd}$-order | 207.8 (8) | 14.1 (5) | 4.0 | 10 |
| | PXRD | Helium | 2$^{nd}$-order | 209.1 (1) | 12.0 (1) | 4.0 | 12 |
| | SCXRD | Helium | 2$^{nd}$-order | unknown | 13.5 (6) | 4.0 | 15 |
| | SCXRD | Helium | 2$^{nd}$-order | unknown | 16.1 (9) | 4.0 | 15 |
| | SCXRD | Silicone oil | 2$^{nd}$-order | 208.4 (8) | 12.4 (6) | 4.0 | a |
| | PXRD | Silicone oil | 2$^{nd}$-order | 208.5 (8) | 19.2 (1) | 4.0 | a |
| $Pmn2_1$ | SCXRD | Silicone oil | 2$^{nd}$-order | 208.5 (5) | 13.0 (3) | 4.0 | a |
| | PXRD | Silicone oil | 2$^{nd}$-order | 208.9 (4) | 19.1 (3) | 4.0 | a |

[a]. This work.

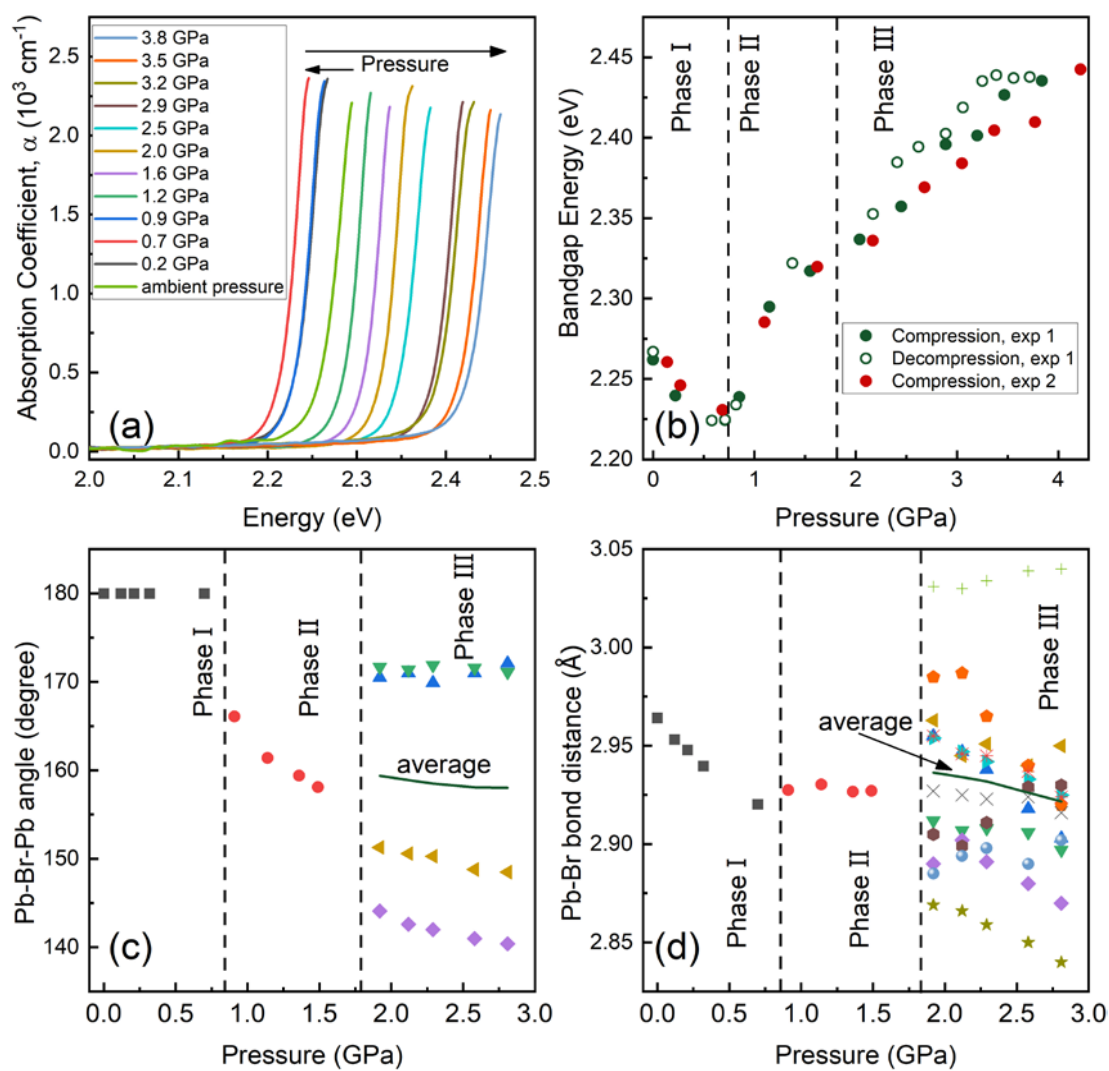

Figure. 5. Pressure dependence of bandgap of MAPbBr$_3$. (a). Optical-absorption spectra of MAPbBr$_3$ at selected pressures from the first experiment (exp 1). (b). Bandgap energy of MAPbBr$_3$ as a function of pressure, the bandgap here at each pressure was derived from the optical absorption spectra shown in (a) by means of a Tauc plot. (c). Pressure dependence of Pb-Br-Pb angles and (d) Pb-Br bond distance obtained from SCXRD experiments. The vertical dash line indicates the phase transition pressure. In figures (c) and (d), the average Pb-Br-Pb angles and Pb-Br bond distance of phase III are shown in solid green lines.

Two independent high-pressure optical-absorption experiments were performed to investigate the bandgap of MAPbBr$_3$. The optical-absorption spectra from the first experiment (exp 1) at selected pressure are shown in **Figure 5a,** and the optical image of the loading at selected pressures can be found in **Figure S1** in the **Supplementary Information** during both the compression and decompression process. The absorption edge first shows a red-shift from room pressure up to around 0.7 GPa, after that the absorption edge exhibits a blue-shift under compression up to 3.8 GPa. We did not conduct any theoretical calculation on the electronic band structure of MAPbBr$_3$, because of the partial occupations in Phase I, but according to the previous calculations,[7,36] the bandgap shows a direct nature. Therefore, the Tauc plot for direct bandgap materials was used to obtain the bandgap energy from the optical-absorption spectra at each pressure,[37] by extrapolating the linear fit of the high-energy part of the ($\alpha h v$)$^2$ vs $hv$ plot to zero, where $\alpha$, $h$ and $v$ are absorption coefficient, Plank constant, and photon frequency, respectively. The bandgap derived from the two optical-absorption experiments (exp 1 and exp 2) are in a good agreement (**Figure 5b**). The bandgap decreases with the increasing pressure in Phase I and increases in Phases II and III with a different slope. The pressure-induced bandgap change is totally reversible, as the bandgap collected at the decompression process of the first experiment shows (**Figure 5b**). According to the previous theoretical calculations, the valence band maximum (VBM) is dominated by the Br-4$p$ orbitals, whilst the conduction band minimum (CBM) is dominated by the Pb-6$p$ orbital.[7] Therefore, the bandgap of MAPbBr$_3$ is strongly affected by the bond distance of Pb-Br and Pb-Br-Pb angle[25]. Furthermore, the positive linear relationship between the Pb-Br bond distance and bandgap energy of MAPbBr$_3$ and MAPbI$_3$ have been established in ref.[8]. On the other hand, the decrease of Pb-Br-Pb angle causes the opening of the bandgap energy in MAPbBr$_3$.

Now the pressure-induced bandgap change of MAPbBr$_3$ can be explained by the pressure dependence of Pb-Br bond distance and Pb-Br-Pb angle as shown in **Figures 5c** and **5d**, which is obtained from SCXRD experiments. In phase I, both Pb and Br atoms are located at only one Wyckoff position (each of them), all the Pb-Br bonds are identical and shortening with increasing pressure, and there is no pressure-induced titling of the PbBr$_6$ octahedra. Therefore, the pressure-induced narrowing of the bandgap energy is caused by the shortening of the Pb-Br bond distance under compression, which favors an increase of atomic hybridization. In Phase II, the Pb-Br bond distance show an independent behavior of pressure, the Pb-Br-Pb angle dramatically bends from 180º to around 165º and further decreases with increasing pressure, so the bandgap starts to broaden under compression. In phase III, Pb is located at two Wyckoff positions and Br is located at eight Wyckoff positions, so there are 12 different Pb-Br bond distances and 4 different Pb-Br-Pb angles. We have calculated the average Pb-Br bond distance and Pb-Br-Pb angles as shown in **Figures 5c** and **5d**. The average Pb-Br slightly decrease with increasing pressure, as well as the average Pb-Br-Pb angles. Those two effects compete under compression, caused a slight increase of the bandgap energy under compression.

## IV. CONCLUSION

In this work, we have reported the results of single-crystal X-ray diffraction (SCXRD), synchrotron-based powder X-ray diffraction (PXRD), and optical-absorption experiments performed on MAPbBr$_3$ perovskite under high pressure. Two pressure-induced phase transition have been independently observed through the three different diagnostics. The crystal structures of each of three MAPbBr$_3$ phases have been determined from high-pressure SCXRD, the transition sequence is $Pm\bar{3}m \rightarrow Im\bar{3} \rightarrow Pmn2_1$, and the phase transitions occurred at 0.8 and 1.8 GPa according to both the SCXRD and PXRD data. The crystal structure determined from SCXRD has been used to perform Rietveld refinements on our PXRD patterns, explaining very well the experiments and supporting the crystal structure determined from SCXRD. The crystal structure in the third phase ($Pmn2_1$) is different from that determined in previous works ($Pnma$)[7,15,16] where only PXRD was used and a full structural determination was not performed.

For each of the three phases, the pressure dependence of the lattice parameters obtained from SCXRD and PXRD, as well as the unit-cell volume per formula unit have been given. The bulk moduli have been calculated by fitting the unit-cell volume data with a second-order Birch-Murnaghan equation of state, the results have been compared with previous works. The bandgap change has been derived from optical-absorption experiments, it shows a narrowing behavior with increasing pressure in Phase I ($Pm\bar{3}m$), while a widening behavior in Phases II ($Im\bar{3}$) and III ($Pmn2_1$), but with a different pressure dependence. There are two effects competing under compression, which results in a nonlinear pressure dependence of the bandgap energy. The pressure-induced shortening of Pb-Br bond distances causes the narrowing of the bandgap energy, while the decrease of the Pb-Br-Pb angles causes the opening of the bandgap energy. The pressure dependence of the Pb-Br bond distance and Pb-Br-Pb angles obtained from SCXRD experiments have been used to explain the bandgap energy change of MAPbBr$_3$ under compression. All the changes found in those three techniques are totally reversible.

# ACKNOWLEDGMENTS

This work was supported by the Generalitat Valenciana under Grant No. PROMETEO CIPROM/2021/075-GREENMAT and by the Spanish Ministerio de Ciencia e Innovación and Agencia Estatal de Investigación (MCIN/AEI/10.13039/501100011033) and the European Union under Grants No. PID2019-106383GB-41/44 and No. RED2018-102612-T (MALTA Consolider Team). A.L. and D.E. thank the Generalitat Valenciana for the Ph.D. Fellowship No. GRISOLIAP/2019/025. R.T. and D.E. thank the Generalitat Valenciana for the postdoctoral Fellowship No. CIAPOS/2021/20. PXRD experiments were performed at the MSPD beamline of ALBA Synchrotron (experiment no. 2021085271.)